\documentclass[a4paper]{article}

\usepackage{INTERSPEECH2022}
\usepackage{multirow}
\usepackage{verbatim}
\usepackage{booktabs}
\usepackage{color}

\usepackage{amsmath, dsfont}
\usepackage{amssymb, bm, mathtools,array}

\newcommand{\PTrial}{$\checkmark$}
\newcommand{\Penroll}{$\times$}

\newcommand{\Nenroll}{$\times$}

\title{Spoofing-Aware Attention based ASV Back-end with Multiple Enrollment Utterances and a Sampling Strategy for the SASV Challenge 2022}
\name{Chang Zeng$^{1,2}$, Lin Zhang$^{1,2}$, Meng Liu$^{3}$, Junichi Yamagishi$^{1,2}$}
\address{
  $^1$National Institute of Informatics, Japan 
  $^2$SOKENDAI, Japan 
  $^3$Tianjin University, China}
\email{\{zengchang,zhanglin,jyamagis\}@nii.ac.jp, liumeng2017@tju.edu.com}

\begin{document}

\maketitle
\begin{abstract}
\vspace{-1mm}
Current state-of-the-art automatic speaker verification (ASV) systems are vulnerable to presentation attacks, and several countermeasures (CMs), which distinguish bona fide trials from spoofing ones, have been explored to protect ASV. However, ASV systems and CMs are generally developed and optimized independently without considering their inter-relationship. In this paper, we propose a new spoofing-aware ASV back-end module that efficiently computes a combined ASV score based on speaker similarity and CM score. In addition to the learnable fusion function of the two scores, the proposed back-end module has two types of attention components, scaled-dot and feed-forward self-attention, so that intra-relationship information of multiple enrollment utterances can also be learned at the same time. Moreover, a new effective trials-sampling strategy is designed for simulating new spoofing-aware verification scenarios introduced in the Spoof-Aware Speaker Verification (SASV) challenge 2022. 
Combining the two types of scores using the proposed back-end optimized by using the sampling strategies, it is confirmed that the SASV-EER can be significantly reduced from 22.91\% to 1.19\% on the evaluation set of the ASVSpoof 2019 LA database.

\end{abstract}
\noindent\textbf{Index Terms}: Speaker verification, Countermeasure, Spoof-Aware Speaker Verification challenge, Attention, Back-end

\section{Introduction}
\vspace{-2mm}


Automatic speaker verification (ASV) aims to determine whether a test utterance was spoken by a particular speaker who was enrolled in advance. Although dramatic improvements have been made in recent years thanks to deep learning, modern ASV systems are vulnerable to several spoofing attacks such as 
text-to-speech (TTS), voice conversion (VC), and replay attacks, and hence countermeasures (CMs) for protecting ASV have been developed \cite{asvspoof2015, asvspoof2017, asvspoof2019, asvspoof2021}. ASV and CMs are active research topics, but the two types of models are currently studied and discussed separately in the research community even though they are related to each other, apart from the integrated ASV and CM metric ``t-DCF'' \cite{t-dcf} and a few prior studies described later. This is the motivation and focus of a new Spoof-Aware Speaker Verification (SASV) challenge 2022 \cite{jung2022sasv}, and our paper also focuses on a new way to make the ASV model itself robust to such spoofing attacks and integrate the two types of models more tightly. 


As mentioned above, there have already been several attempts on this topic. For instance, multi-task learning based optimization of ASV and CM models was reported in \cite{multitask, Zhao2022mtl-sasv}. In \cite{anssi2022_RL_asvcm}, a reinforcement learning based approach is used instead in order to directly optimize both ASV and CM models with respect to the non-differentiable t-DCF metric. In addition, back-propagation based joint training of neural probabilistic linear discriminant analysis (NPLDA) \cite{ramoji2020nplda} and CM using light convolutional neural networks (LCNN) \cite{lavrentyeva2019stc,wang2021comparative} using an approximated differentiable t-DCF loss was also studied \cite{anssi2022_RL_asvcm}. There have also been attempts to combine the two types of models at the embedding level \cite{joint-iv, joint-op} or at the score level \cite{integrated}.

Unlike prior studies, this paper focuses on learnable fusion of an ASV score and CM score as a part of a neural network based ASV back-end module.  This approach is based on our previous work \cite{attention-backend} in which we proposed an attention-based back-end model that learns the intra-relationships of the enrollment utterances for ASV. The back-end module merges multiple neural speaker embeddings extracted from each enrollment utterance of a registered target speaker by means of two stacked attention networks to compute a speaker representative vector that is more suitable than simply averaging the embeddings. This network can be directly optimized based on back-propagation using training data containing various positive and negative pairs. Here, the negative pairs correspond to cases where the speaker ID of enrollment utterances and that of a test trial does not match. We have shown that this attention back-end is effective for multiple speaker encoders, including a self-attention based speaker encoder \cite{tdnn-asp} and the emphasized channel attention, propagation and aggregation in a time delay neural network (ECAPA-TDNN) speaker encoder \cite{ecapa-tdnn}.

To make the above ASV back-end robust against spoofing attacks and to evaluate it in the SASV Challenge 2022, two new improvements are proposed in this paper. The first improvement is to use CM scores in the neural ASV backend through score fusion. Score fusion itself is not new, but since the score fusion is conducted inside the neural network for the ASV back-end, gradients computed through the fused score are affected by the CM score. In other words, even a simple fusion results in an update of the ASV backend parameters that takes into account both the CM score and its impact. 
The second improvement is a new definition of positive and negative pairs for trials that accounts for spoofed data, such as those introduced in the SASV challenge 2022. Unlike our previous experiments, we also define  spoofed test trials using TTS or VC as an additional type of negative cases, regardless of their speaker similarity. 
We describe how to efficiently create and sample such positive and negative cases from a mini-batch during the training phase.

The rest of this paper is organized as follows: The proposed spoofing-aware attention back-end with the novel trials-sampling strategy is explained in Section \ref{sec:SASB}. Experimental conditions are described in Section \ref{sec:exp_res}, and results are shown in Section \ref{sec:res}. This paper is concluded in Section \ref{sec:conclusion}.

\begin{figure*}[!t]
\centering
\includegraphics[width=1.35\columnwidth]{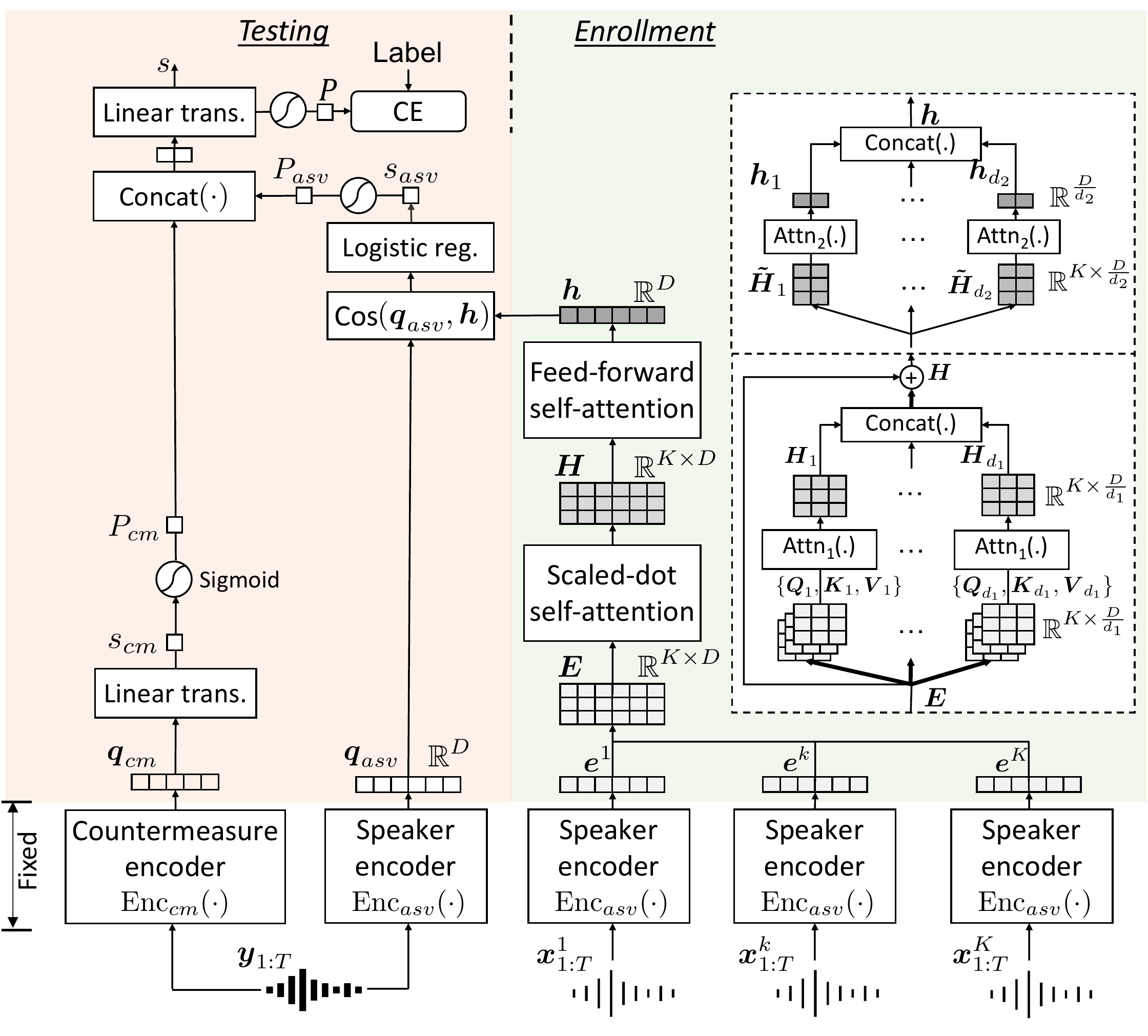}
\vspace{-3mm}
\caption{Architecture of the proposed spoofing-aware attention back-end including score-level fusion.}
\label{fig:attention-backend}
\vspace{-3mm}
\end{figure*}

\section{Spoofing-aware attention back-end}
\label{sec:SASB}

\subsection{Model architecture}
The network architecture of the proposed and extended attention back-end model is shown in Figure \ref{fig:attention-backend}. Suppose an enrolled speaker has $K$ enrollment utterances $\{\boldsymbol{x}_{1:T}^{1}, \cdots, \boldsymbol{x}_{1:T}^{K}\}$. Enrollment embedding vectors $\{\boldsymbol{e}^{1}, \cdots, \boldsymbol{e}^{K}\}$ are extracted from each of the enrollment utterances using a pre-trained speaker encoder $\rm{Enc}_{asv}(\cdot)$ and then they are further converted to a speaker representative vector $\boldsymbol{h}$ per enrolled speaker through scaled-dot self-attention (SDSA) \cite{Vaswani2017-Attention} and feed-forward self-attention (FFSA) \cite{Raffel2015-FFNATT,Lin2017-StructureATT} modules. The role of the attention networks is to learn the intra-relationships of the multiple enrollment speaker embeddings. For details of the attention networks, please see \cite{attention-backend}. 

At the test phase, a speaker embedding $\boldsymbol{q}_{asv}$ and CM embedding $\boldsymbol{q}_{cm}$ are first extracted from a test trial $\boldsymbol{y}_{1:T}$ using pre-trained encoders. The  speaker encoder $\rm{Enc}_{asv}(\cdot)$ is the same as that used for the enrollment process. The CM encoder $\rm{Enc}_{cm}(\cdot)$ is also a pre-trained network. Using the two embeddings $\boldsymbol{q}_{cm}$ and $\boldsymbol{q}_{asv}$ of the test trial as well as the speaker representative vector $\boldsymbol{h}$ of a claimed speaker, the proposed back-end module produces a joint probability score $s$ based on two different hypotheses.





The first one is the CM's hypothesis $H^{cm}$ that determines the utterance $\boldsymbol{y}_{1:T}$ is bona fide, which is modeled by a simple sigmoid function:
\begin{equation}
    \label{equa:Pcm}
    H^{cm}: P_{cm}(s_{cm})= \frac{1}{1 + \exp^{-s_{cm}}},
\end{equation}
where $P_{cm}$ is the CM probability that represents how likely the input is bona fide. And the CM score $s_{cm}$ is obtained via a linear transformation of the CM embedding $\boldsymbol{q}_{cm}$ as shown in Fig.\ \ref{fig:attention-backend}. 

The second one is the ASV hypothesis $H^{asv}$, which indicates that $\boldsymbol{y}_{1:T}$ is uttered by the speaker who enrolled his or her identity with multiple utterances. 
%
%
The probability of $H^{asv}$ is calculated based on cosine distance of $\boldsymbol{q}_{asv}$ and $\boldsymbol{h}$, followed by a sigmoid function below:  
%
\begin{align}
    \begin{split}
    \label{equa:Pasv}
    H^{asv}: P_{asv}(\boldsymbol{q}_{asv},\boldsymbol{h}) 
    &= \frac{1}{1 + \exp^{-s_{asv}}} \\
    &= \frac{1}{1 + \exp^{-a \, \text{Cos}(\boldsymbol{q}_{asv},\boldsymbol{h}) - b}}, 
    \end{split}\\
    \begin{split}
    \text{Cos}(\boldsymbol{q}_{asv},\boldsymbol{h}) &=\frac{\boldsymbol{q}_{asv} \cdot \boldsymbol{h}}{||\boldsymbol{q}_{asv}||\cdot ||\boldsymbol{h}||},
    \end{split}
\end{align}
where $P_{asv}(\boldsymbol{q}_{asv},\boldsymbol{h})$ denotes the probability of $\boldsymbol{q}_{asv}$ and $\boldsymbol{h}$ belonging to the same speaker, 
and $a$ and $b$ are trainable parameters for the calibration.

The CM probability $P_{cm}$ and the ASV probability $P_{asv}$ are concatenated as a 2-dimensional vector, which is then projected to the final score $s$ by a linear transformation, 
\begin{align}
    \begin{split}
        P(P_{cm},P_{asv}) & = \frac{1}{1+\exp^{-s}} \\
        & = \frac{1}{1+\exp^{-(w_1*P_{cm}+w_2*P_{asv}+v)}},        
    \end{split}
    \label{eq:fusion}
\end{align}
%
where $P(P_{cm}, P_{asv})$ denotes the probability of a joint decision using CM and ASV, which will be used for computing the cross-entropy (CE) loss,
and $w_1$, $w_2$ and $v$ are trainable parameters in the linear transformation.

These formulations, including the fusion of the CM and the ASV probabilities in Eq.\ \ref{eq:fusion}, are standard ones, but a scientifically interesting point is that gradients computed through the fusion are affected by the CM probability. Hence, the updated ASV backend parameters (e.g.\ attention weights of SDSA and FFSA) become CM aware\footnote{Then, the linear transformation matrix for the CM embedding vector becomes ASV-aware.} and so our ASV back-end is expected to take into account the CM impact and hopefully become more robust to spoofing attacks.

\begin{table}[t]
\footnotesize
    \centering
    \caption{Composition of triples of (test-CM-embedding, test-speaker-embedding, enrollment-data) for training back-end model and ground-truth labels from mini-batch. A, B, and C are speaker IDs, and 1, 2, 3 and 4 are his or her audio index and ones without underline indicate spoofed audio using TTS or VC. \PTrial and \Penroll denote test and enrollment audio files, respectively.  \label{table:train}}
    \vspace{-2mm} 
    \setlength{\tabcolsep}{4pt}
    {
    \begin{tabular}{cccccccccccccc}
 \cmidrule{2-13} 
 & \multicolumn{4}{c}{$A$} & \multicolumn{4}{c}{$B$} & \multicolumn{4}{c}{$C$} \\
 \cmidrule{2-13} 
 & \underline{1} & 2 & \underline{3} & 4 & 1 & \underline{2} & 3 & \underline{4} & 1 & 2 & \underline{3} & \underline{4} & Label\\
 \cmidrule{2-13}
 \multirow{10}*{\rotatebox[origin=c]{90}{\textbf{Pairs to be used for training}}} & \PTrial & \Penroll & \Penroll & \Penroll & & & & & & & & & P \\
 
 & \PTrial &  &  &  &  & \Nenroll  & \Nenroll & \Nenroll  & &  & & & N \\
 
 & \PTrial &  &  &  &  & &  & & & \Nenroll  & \Nenroll & \Nenroll &  N \\

 & \Penroll & \PTrial & \Penroll & \Penroll & & & & & & & & & N \\
 
 & & \PTrial &  &  & \Nenroll & & \Nenroll & \Nenroll &  &  & & & N \\
 
 & & \PTrial &  &  &  &  &  & & \Nenroll & & \Nenroll & \Nenroll & N \\

 
 
 
 & & & & &  & & \vdots & & & & &  & \\

 & \Nenroll & \Nenroll & \Nenroll & &  &  &  &  &  &  & & \PTrial  & N \\
 
 & & & &  & \Nenroll & \Nenroll & \Nenroll &   &  &  &  & \PTrial & N \\
 
 & &  &  &   & & & & & \Penroll & \Penroll & \Penroll & \PTrial & P \\
 \cmidrule{2-13}
\end{tabular}
}
\vspace{-2mm}
\end{table}

\subsection{Sampling strategy adopted for the SASV challenge}

\label{sec:sampling}

Next, we explain how the positive and negative pairs are defined and how they are sampled from multiple audio files included in a mini-batch so we can train the proposed neural back-end with the CE criteria. In our previous study \cite{attention-backend}, positive and negative pairs were defined in the absence of spoofed speech using TTS or VC. In other words, we consider zero-effort impostors (non-target speakers) only and the negative pairs correspond to cases where speaker IDs of enrollment utterances and the test trial do not match. In this paper, in order to make the back-end spoofing-aware, spoofed test trials using TTS or VC are also defined as negative, even if the speaker identity of the spoofed speech matches the target speaker accurately.

For each mini-batch, assuming it has $M$ speakers, and that each speaker has $K$ audio files, the size of one mini-batch is $M\times K$. 
Considering that the number of bona fide audio samples is limited in the training dataset, the number of bona fide and spoofed audio samples for each speaker in one mini-batch is set to be equal, which means that one speaker in a mini-batch has $K/2$ bona fide and $K/2$ spoofed audio files. In addition, 
speaker embeddings extracted from $K/2$ bona fide audio files are rearranged to form speaker verification trials which have multiple enrollment utterances.  Table~\ref{table:train} illustrates one example to form the positive and negative pairs. In this example, one mini-batch has $M=3$ speakers $A$, $B$, and $C$, and each speaker has $K=4$ audio files consisting of 2 bona fide audio files and 2 spoofed audio files whose index ranges from 1 to 4, and ones without underline indicate spoofed samples using TTS or VC. \textit{When a spoofed sample is selected as enrollment, we mask it as a zero vector \cite{Vaswani2017-Attention} to exclude from training.}


There are numerous ways to compose the positive and negative cases. For the SASV challenge, we adopted the following: First, one audio file of one test speaker is selected as a test trial from the mini-batch, and the rest of his or her audio files are preserved as enrollment utterances. This results in pairs where speaker IDs are matched, but the audio files may or may not be spoofed by TTS or VC. Therefore, in these pairs, if the test trial is spoofed by TTS or VC, the label is set to N (negative), otherwise it is set to P (positive). Next, we consider pairs where the speaker IDs do not match, by simply combining the test trial of one speaker with the audio files of another speaker in the given mini-batch. In this case, the label is set to N. By considering the logical sum of the ASV and CM like the above table, we can define various positive and negative cases from the audio files included in the mini-batch.

\subsection{Loss calculation using hard training samples}
\label{sec:loss}

As described above, the proposed back-end network is updated using a set of the positive and negative pairs based on the binary CE criterion. Note that the sampling strategy in the previous section results in a large difference in the amount of positive and negative pairs, which can create a bias towards the negative class, so we devise a way to deal with this.
%
%
Specifically, to alleviate the impact of unbalanced classes and emphasize the contribution of hard training samples \cite{bd-lmcl}, we select all positive samples and use only the top $N$ most difficult negative samples when we compute the CE loss for backward propagation.

\section{SASV Challenge Results}

\label{sec:exp_res}
\subsection{Datasets and pre-trained ASV and CM}

We conducted experiments on the ASVspoof2019 LA \cite{Wang2020data} dataset using the SASV protocol. There are 20, 20, and 67 speakers in the train, dev, and eval sets, respectively. The training data contains total of 25,380 utterances from 20 speakers, the dev set contains 24,844 utterances from different 20 speakers, and the eval set includes 71,237 utterances from 67 different speakers. Since our proposed model focuses on integrating ASV and CM information from a back-end view, we utilized pre-trained ECAPA-TDNN \cite{desplanques2020ecapa} and Audio Anti-Spoofing using Integrated Spectro-Temporal (AASIST) Graph Attention Networks \cite{Jung2022AASIST} models for extracting speaker and CM embeddings, respectively. 






\subsection{Training methodology}
In our experiments, one mini-batch contains 16 speakers and each speaker has 5 bona fide and 5 spoofed audio samples. In order to optimize the proposed back-end model, the SGD optimizer with 0.0001 learning rate, 0.9 momentum and 0.00001 weight decay was utilized to update the model for 40 epochs. The learning rate is decayed by 0.95 at the end of each epoch. In addition, as described in Section \ref{sec:loss}, only the top 100 trials with the largest loss values were selected for backward propagation. The speaker encoder and countermeasure were frozen during the back-end training.

\subsection{Evaluation metrics}
The SASV-EER in percentage was used as the primary metric for performance assessment. This metric mixes non-target and spoofing impostor as the negative class \cite{jung2022sasv}. Besides, according to the SASV evaluation plan, SV- and SPF-EERs were also used to evaluate the performance of ASV on non-target and CM on spoofing, respectively.

\subsection{SASV reference and baseline systems}

An ASV reference system and two baseline systems were provided by the SASV challenge 2022 \cite{jung2022sasv}. The ASV reference system is based on a pre-trained ECAPA-TDNN model \cite{desplanques2020ecapa}  and does not use any CM information. The two baseline systems were based on the reference ASV system and a pre-trained AASIST \cite{Jung2022AASIST} model and they are as follows:

\begin{description}
\item[ASV-CM score sum:] The ASV-CM score sum method simply sums scores generated by the separate pre-trained ASV and CM models. Thus, no data is used for this baseline as it does not involve any training or fine-tuning.
\item[ASV-CM embedding fusion:] For ASV-CM embedding fusion, embeddings extracted from the pretrained ASV and CM models are fused via a simple feedforward network with three hidden layers, trained using the ASVspoof 2019 LA train partition. See \cite{jung2022sasv} for details.
\end{description}

\subsection{Challenge Results}

\begin{table}[tb]
\caption{
  EERs (\%) for the SASV 2022 development and evaluation partitions.
  SASV-EER for all systems are calculated using the entire protocol that includes trials used to measure the SV-EER (target vs.\ non-target) and those used to measure the SPF-EER (target vs.\ spoof).
  Results are shown for a reference ASV system (ECAPA-TDNN), the two baseline systems, and our attention back-end with and without the CM modules.\label{table:results}}
\vspace{-4mm}  
\setlength{\tabcolsep}{2pt}
\footnotesize
\begin{center}
\begin{tabular}{lcccccc}
\toprule
\multirow{2}{*}{} & \multicolumn{2}{c}{SV-EER[\%]}  & \multicolumn{2}{c}{SPF-EER[\%]} & \multicolumn{2}{c}{SASV-EER[\%]}\\ 
\cline{2-7}
                  & \multicolumn{1}{c}{Dev} & \multicolumn{1}{c}{Eval} &  \multicolumn{1}{c}{Dev} & \multicolumn{1}{c}{Eval}& \multicolumn{1}{c}{Dev} & \multicolumn{1}{c}{Eval}               \\ 
\hline
\textbf{SASV baselines \cite {jung2022sasv}} \\
ASV only          & \phantom{0}1.88  & \phantom{0}1.63  & 20.30 & 30.75 & 17.38  & 23.83 \\
ASV-CM score sum & 32.88 & 35.32 & \phantom{0}0.06  & \phantom{0}0.67  & 13.07  &19.31  \\
ASV-CM embed.\ fusion  & 12.87 & 11.48 & \phantom{0}0.13  & \phantom{0}0.78  & \phantom{0}4.85   & \phantom{0}6.37   \\
\midrule 
\textbf{Attention back-end} \\
ASV only \cite{attention-backend} & \phantom{0}1.54  & \phantom{0}1.42  & 19.81 & 29.62 & 16.78  & 22.91 \\
Our method            & \phantom{0}1.41 & \phantom{0}1.32 & \phantom{0}0.61 & \phantom{0}1.14 & \phantom{0}0.81 & \phantom{0}1.19 \\
\bottomrule
\end{tabular}
\end{center}
\vspace{-7mm}
\end{table}

Table \ref{table:results} shows SV-, SPF-, and SASV-EERs in percentage of the reference, baseline systems and our systems based on attention based back-end. For reference, results of  our attention back-end  without the CM modules are also included\footnote{This system was trained according to \cite{attention-backend} except for the mini-batch size. Only 16 speakers were sampled in each mini-batch due to the limitation of the number of speakers in the training dataset}. 

Comparing SV-EER with SASV-EER of the first row of the table, we can first see how vulnerable the ECAPA-TDNN-based ASV model is when attacked by spoofed data. The EER of ECAPA-TDNN on the eval set was severely degraded from 1.63\% to 23.83\%. Next, observing the result of the ASV-CM score sum shown in the second row, introducing a CM score at the inference stage and summing the ASV and CM scores can alleviate the vulnerability of the ASV model to some extent. However, due to there being no learnable parameters in this baseline model, one can see that the result is still unacceptable. Then, as expected, ASV-CM embedding fusion using the simple feedforward network led to large improvements in terms of SASV-EER as shown in the third row of the table. 

Next, we focus on results on our attention back-end. From the fourth row of the table, we can see that our attention back-end has slightly better SV-EERs than the SASV reference ASV system even though they use the same ECAPA-TDNN based speaker encoder. This is due to its attention networks that merge multiple enrollment utterances in a nonlinear way, and is consistent with our previous results \cite{attention-backend}. At the same time, however, the SASV-EER results also indicate that our ASV backend is also vulnerable to spoofing attacks.

Finally, the fifth row of the table clearly shows that our extended attention back-end became robust to the spoofing attacks while preserving the original ASV performance. Its SASV-EER on the eval set was reduced from 22.91\% to 1.19\%. Its SV-EERs on the eval set are almost the same or even slightly better compared to its counterpart result of our attention back-end without the CM module (1.32\% vs. 1.42\%). These results are better than those of the SASV baselines apart from SPF-EER results. 



\section{Post-challenge Analysis}
\label{sec:res}

Our backend has the ability to fuse the ASV and CM scores and to merge speaker embeddings extracted from multiple enrollment utterances, both of which are learned from the training set in the ASVspoof database including spoofed audio. The ability to merge the ASV and CM scores is essential and cannot be excluded due to the nature of the SASV challenge, but in this section we analyze whether there is a benefit to learning the attention based module to merge enrollment utterances together as well.  To answer this question, we have additionally created a system in which the above attention module is replaced by a simple averaging operation of the embedding vectors extracted from the enrollment utterances, excluding the learnable weights. 
The second row of Table \ref{table:ablation} shows results for the system where the learnable weights for multiple enrollment utterances were excluded and hence no training process was conducted for creating the speaker representative vector $\boldsymbol{h}$. Interestingly, this system improved the SPF-EER results. By excluding the learnable parameters of the attention module for enrollment utterances, the influence of CM on the final score seems to increase.  Conversely, however, the SV-EER and SASV-EER results deteriorated compared to our system using both of them. It appears that the attention module for the enrollment utterances is also necessary to maintain good ASV performance and robustness against spoofing attacks.

\begin{table}[t]
\caption{
  EERs (\%) for post-challenge analysis of the proposed method. A system ``w/o attention'' corresponds to one in which the attention modules were replaced to a simply averaging operation of multiple enrollment speaker embeddings.}
\label{table:ablation}
\vspace{-4mm}
\setlength{\tabcolsep}{3pt}
\footnotesize
\begin{center}
\begin{tabular}{lcccccc}
\toprule
\multirow{2}{*}{} & \multicolumn{2}{c}{SV-EER[\%]}  & \multicolumn{2}{c}{SPF-EER[\%]} & \multicolumn{2}{c}{SASV-EER[\%]}\\ 
\cline{2-3} \cline{4-5} \cline{6-7}
                  & \multicolumn{1}{c}{Dev} & \multicolumn{1}{c}{Eval} &  \multicolumn{1}{c}{Dev} & \multicolumn{1}{c}{Eval}& \multicolumn{1}{c}{Dev} & \multicolumn{1}{c}{Eval}               \\ 
\hline
Our method                            & 1.41 & 1.32 & 0.61 & 1.14 & 0.81 & 1.19 \\
\phantom{O}w/o attention                         & 2.09 & 1.97 & 0.07 & 0.76 & 1.15 & 1.53 \\

\bottomrule
\end{tabular}
\end{center}
\vspace{-7mm}
\end{table}

\section{Conclusions}
\label{sec:conclusion}

In this paper, we have proposed a new spoofing-aware ASV back-end module that efficiently combines an ASV score based on speaker similarity and CM score. It also contains the attention components that merge multiple enrollment utterances into the speaker representative vector. A new sampling strategy was also designed for SASV challenge 2022. The SASV challenge results have proved that combining the two types of scores using the proposed back-end optimized by using the sampling strategies, our extended attention back-end became robust to spoofing attacks while preserving the original ASV performance. The SASV-EER was reduced from 22.91\% to 1.19\% on the eval set. We also found that the score fusion and attention modules are complementary.

\vspace{1mm}
\noindent
\textbf{Acknowledgements:} We would like to thank Dr.\ Xin Wang and Dr.\ Erica Cooper for their comments. We would like to thank the organizers of the SASV challenge 2022. This study is partially supported by JST CREST Grants (JPMJCR18A6, JPMJCR20D3 and JPMJFS2136) and by MEXT KAKENHI Grant (21H04906).

\bibliographystyle{IEEEtran}

\bibliography{mybib}

\end{document}